\documentclass[twocolumn,showpacs,showkeys,preprintnumbers,amsmath,amssymb]{revtex4}

\usepackage{graphicx}
\usepackage{dcolumn}
\usepackage{bm}

\begin{document}

\preprint{APS/123-QED}

\title{Cross section measurements of $\alpha$-induced reactions on $^{92,94}$Mo and $^{112}$Sn for $p$-process studies}

\author{W. Rapp}
\altaffiliation[Present address: ]{Westinghouse Electric Germany GmbH, Dudenstrasse 44, D-68167 Mannheim.}

\author{I. Dillmann}
 \email{iris.dillmann@ph.tum.de}
 \altaffiliation[Present address: ]{Physik Department E12, Technische Universt\"at M\"unchen, D-85748 Garching}
\author{F. K\"appeler}
\affiliation{Institut f\"ur Kernphysik, Forschungszentrum Karlsruhe, Postfach 3640, D-76021 Karlsruhe, Germany}

\author{U. Giesen}
\author{H. Klein}
\affiliation{Physikalisch-Technische Bundesanstalt, Bundesallee 100, D-38116 Braunschweig, Germany}

\author{T. Rauscher}
\affiliation{Departement f{\"u}r Physik, Universit{\"a}t Basel, Klingelbergstrasse 82, CH-4056 Basel}

\author{D. Hentschel}
\author{S. Hilpp}
\affiliation{Institut f\"ur Nukleare Entsorgung, Forschungszentrum Karlsruhe, Postfach 3640, D-76021 Karlsruhe, Germany}

\date{\today}

\begin{abstract}
The $^{92}$Mo($\alpha,n$)$^{95}$Ru, $^{94}$Mo($\alpha,n$)$^{97}$Ru, and $^{112}$Sn($\alpha,\gamma$)$^{116}$Te
cross sections were measured at the upper end of the $p$-process Gamow window between 8.2 MeV and 11.1 MeV. Our results are slightly lower than global Hauser-Feshbach calculations from the code NON-SMOKER, but still within the uncertainty of the prediction. The $^{112}$Sn($\alpha,\gamma$)$^{116}$Te cross section agrees well with a recently measured thick-target cross section in the same energy range. For the $^{92,94}$Mo($\alpha,n$) reactions the present data close to the reaction thresholds could eliminate previous uncertainties within a factor of 20, and we can present now useful comparisons to statistical model calculations with different optical potentials.
\end{abstract}


\maketitle

\section{Introduction}
A ``$p$ process'' was postulated to produce 35 stable but rare isotopes between $^{74}$Se and $^{196}$Hg on the proton-rich side of the valley of stability. Unlike the remaining 99\% of the heavy elements beyond iron these isotopes cannot be created by (slow or rapid) neutron captures \cite{bbfh57}, and their solar and isotopic abundances are 1-2 orders of magnitude lower than the respective $s$- and $r$-process nuclei \cite{AnG89,iupac}. However, so far it seems to be impossible to reproduce the solar abundances of all $p$ isotopes by one single process. In current understanding, several (independently operating) processes seem to contribute.

The largest fraction of $p$ isotopes is created in the ``$\gamma$ process'' by sequences of photodissociations and
$\beta^+$ decays \cite{woho78,woho90,ray90,argo03}. This occurs in explosive O/Ne burning during SNII explosions and reproduces the solar abundances for the bulk of $p$ isotopes within a factor of $\approx$3 \cite{ray90,raar95}. The SN shock wave induces temperatures of 2-3 GK in the outer (C, Ne, O) layers, sufficient for triggering the required photodisintegrations. More massive stellar models (M$\geq$20~M$_\odot$) seem to reach the required temperatures for efficient photodisintegration already at the end of hydrostatic O/Ne burning \cite{rahe02}. Historically, the $p$ process was thought to proceed via proton captures but today they are found to play no role since the required amount of free protons is not available in the Ne and O layers within the $p$-process timescale of a few seconds and proton captures are too slow for elements with large $Z$. Therefore, the term "$\gamma$ process" is also often used synonymously because seed nuclei pre-existing in the stellar plasma are photodissociated and more proton-rich isotopes are produced, initially by $(\gamma,~n)$ reactions. When ($\gamma,~p$) and ($\gamma,~\alpha$) reactions become comparable or faster than neutron emission within an isotopic chain, the reaction path branches out and feeds nuclei with lower charge number $Z$.
The decrease in temperature after passage of the shock leads to a freeze-out via neutron captures and mainly $\beta^+$ decays, resulting in the typical $p$-process abundance pattern with maxima at $^{92}$Mo ($N$=50) and $^{144}$Sm ($N$=82).

However, the $\gamma$-process scenario suffers from a strong underproduction of the most abundant $p$ isotopes, $^{92,94}$Mo and $^{96,98}$Ru, due to lack of seed nuclei with $A$$>$90. For these missing abundances, alternative processes and sites have been proposed, either related to strong neutrino fluxes in the deepest ejected layers of a SNII ($\nu p$ process \cite{FML06}), or rapid proton-captures in proton-rich, hot matter accreted on the surface of a neutron star ($rp$ process \cite{scha98,scha01}). 

Modern, self-consistent studies of the $\gamma$ process have problems to synthesize $p$ nuclei in the regions $A<124$ and $150\leq A\leq 165$ \cite{rahe02}. It is not yet clear whether the observed underproductions are only due to a problem with astrophysical models or also with the nuclear physics input, i.e. the reaction rates used. Thus, the reduction of uncertainties in nuclear data is strictly necessary for a consistent understanding of the $p$ process. Experimental data can improve the situation in two ways, either by directly replacing predictions with measured cross sections in the relevant energy range or by testing the reliability of predictions at other energies when the relevant energy range is not experimentally accessible. 

Accordingly, model calculations of the $p$ process require large nuclear reaction networks involving about 1800 isotopes and more than ten thousand reaction rates. Since the largest fraction of such a network contains proton-rich, unstable isotopes, which are not accessible for cross section measurements with present experimental techniques, almost all of these rates have to be determined theoretically by means of the statistical Hauser-Feshbach (HF) formalism \cite{hafe52}. Hence, there is no alternative to employing statistical model calculations, e.g. those
performed with the codes NON-SMOKER \cite{rath00,rath01} or MOST \cite{most05}. Comparison with experimental results has shown that theoretical HF predictions with global parameter sets reproduce the neutron- and proton-induced cross sections within a factor of two or even better. This deviation increases for $\alpha$-induced reactions. 
It has been noted that the difficulties in calculating these rates are due to the poor knowledge of the $\alpha$-optical potential at astrophysically relevant energies \cite{MRO97,DGG02}. The optical potential governs the transmission coefficient in the HF model. Additional input to this model are nuclear level densities and particle separation energies. The latter data are well defined since nuclear masses in the $p$-process network are experimentally known.

It is only one decade ago that charged particle-induced reaction rates at $p$-process energies ($E_p$$\approx$1-10~MeV, $E_\alpha$$\approx$5-15~MeV) became subject of experimental efforts (for a list of reactions, see \cite{kadonis}).  For example, a series of $\alpha$-scattering  experiments at energies below the Coulomb barrier helped to improve the optical $\alpha$-nucleus potential \cite{DGG02, AOP03, FGM03, Rau03a, Rau03b}. However, firm conclusions are still difficult to draw due to the lack of experimental cross sections.

Nevertheless, extrapolations to the low astrophysical energies remain necessary since the Coulomb barrier hampers scattering experiments in the astrophysically relevant energy range. Scattering data within a confined energy range can be described 
equally well by different optical potentials with different extrapolation behavior. This ambiguity, which is called the family problem, makes extrapolations very uncertain \cite{McS66,fuki96,MRO97,GFG05}. These uncertainties can be avoided if experimental cross sections are available for comparison with HF predictions in or close to the astrophysical Gamow window. 

Up to now only five $(\alpha,\gamma)$ measurements with relevance for the $p$ process were performed within the Gamow window between 5 and 15~MeV: $^{70}$Ge \cite{fuki96}, $^{96}$Ru \cite{rapp02}, $^{106}$Cd \cite{GKE06}, $^{112}$Sn \cite{OEG07}, and $^{144}$Sm \cite{somo98a}. With exception of the latter, the first four experiments agree within a factor of three with the NON-SMOKER prediction using a global parameter set with the McFadden-Satchler $\alpha$ potential \cite{McS66}. For the $(\alpha,n)$ particle-exchange reactions a much larger amount of data exists, but almost all of these measurements are restricted to the upper end of the Gamow window. It was noted in \cite{RGW06,Rau06} that especially in the heavy mass region (beyond $N$=82) the knowledge of $(\alpha,x)$ cross sections is crucial, whereas at lower masses the $p$-induced reactions have more importance. However, measurements of $\alpha$-induced reactions are mainly limited -- apart from some exceptions for $(\alpha,n)$ cross sections -- to the lower mass region below $^{144}$Sm. In an attempt to constrain the $\alpha$+nucleus optical potential at low energy, it has been suggested to use ($n,\alpha$) reactions and some reactions around $A\approx 147-152$ have been studied in this way \cite{GKA00,KGA01,KGR04}.

Finally, it has to be emphasized that with a few exceptions ($\alpha,~\gamma$) and ($p,~\gamma$) data in the relevant energy range are more useful for a direct application in reaction networks than ($\gamma,~\alpha$) or ($\gamma,~p$) data despite of the fact that the $p$ process proceeds via photodisintegration reactions. This may seem puzzling at first sight but has two reasons. First, reactions in the stellar plasma involve thermally excited targets and not only targets in the ground state as in the laboratory. Measured rates have to be corrected for this effect. It can be shown that the thermal effects are much more pronounced in the photodisintegration than in the capture channel. Thus the measured rate is closer to the actual rate when considering capture reactions. The reverse rate can then be derived applying detailed balance. Second, in order to avoid inconsistencies, forward and reverse rates for a given reaction in a reaction network have to stem from the same source, related to each other by detailed balance. Because of the exponential factor $\exp (-Q_\mathrm{forward})$ in the expression for the relation of reverse to forward rate (see, e.g., \cite{rath00}), it is preferrable to have data for the direction of positive $Q$ value to limit numerical errors in the simulation. In most cases this is the capture direction, exclusively in the case of proton captures close to stability. However, in the intermediate to heavy mass region the $p$ process also includes $\alpha$ emitters with negative $\alpha$ capture $Q$ values.
 
In this work we study ($\alpha,n$) reactions on light Mo isotopes and $\alpha$ capture on $^{112}$Sn, motivated by the need to better understand the above described deficiencies in the $p$ process models. Previous investigations of astrophysical relevance were restricted to $\alpha$ scattering on $^{92}$Mo \cite{FGM03}, $^{106}$Cd \cite{KFG06}, and on $^{112}$Sn \cite{GFG05} attempting to derive an improved optical potential for these cases. The experimental setup for the measurement of $\alpha$-induced cross sections is described in Sec.~\ref{exp}. The resulting cross sections and a comparison with previous experimental results for $^{92}$Mo($\alpha,n$)$^{95}$Ru, $^{94}$Mo($\alpha,n$)$^{97}$Ru, and $^{112}$Sn($\alpha,\gamma$)$^{116}$Te are given in Sec.~\ref{results}. The impact of using different optical $\alpha$+nucleus potentials in theoretical calculations for these reactions is investigated in Sec.~\ref{sec:alphapot}.

\section{Cross section measurements}\label{exp}

All cross section measurements have been carried out at the cyclotron of the Physikalisch-Technische Bundesanstalt (PTB) in Braunschweig/ Germany \cite{BCD80} with the activation technique by irradiation of thin sample layers and subsequent counting of the induced activity. A detailed description of the measurement technique and the data analysis can be found in Ref.~\cite{rapp02}.  

\subsection{Preparation and characterization of samples}
The sample material was deposited on 0.4~mm thick Ta backings, which were thick enough to stop the $\alpha$-beam completely to ensure a reliable charge collection. Background reactions are efficiently suppressed by the high Coulomb barrier of Ta because $\alpha$ capture on $^{181}$Ta leads to the stable product nucleus $^{185}$Re. The $^{181}$Ta$(\alpha,n)$$^{184}$Re channel opens at 9.86~MeV but does not interfere with our measurements. The contribution from the extremely rare isotope $^{180}$Ta$^{\rm m}$ (natural elemental abundance 0.012\% \cite{iupac}) could be neglected. 
  
The Mo samples were produced by sputtering natural metallic Mo in a pure Ar atmosphere. These samples were 10 mm in diameter and exhibited an area density between 5$\times$10$^{17}$ and 3$\times$10$^{18}$ Mo atoms/cm$^2$, corresponding to layer thicknesses between 100~nm and 600~nm. The $^{112}$Sn samples were prepared from 10 mg of enriched metal powder containing 98.9\% of $^{112}$Sn. This batch was dissolved in 30\% hydrochloric acid and diluted with 90\% acetone. The solution was then deposited on the Ta backings, where uniform SnCl$_2$ layers with an area density between 6.2 and 9.8$\times$10$^{17}$ Sn atoms/cm$^2$ (layer thicknesses $\approx$300-450~nm) were obtained after evaporation of the acetone. The adhesion of the layers was improved by soft sand-blasting of the Ta backings before the deposition. 
   
The masses of the Mo and Sn samples were determined by X-ray fluorescence analysis using a crystal spectrometer (Siemens SRS 3000), which was operated with a LiF (100) crystal and a rhodium anode. The fluorescence yield was measured with a gas counter and a NaI detector. The efficiency of the spectrometer was calibrated by the radiation emitted from an empty backing and by nine well defined Mo and Sn reference samples. The latter were
prepared from standard solutions of (NH$_4$)$_2$MoO$_4$ and SnCl$_2$, respectively. The sample mass was 
derived from the measured K$_{\alpha 1}$ and K$_{\beta1}$ X-ray lines. In case of the Sn sample an additional Ta diaphragm 10 mm in diameter was placed on top of the sample to restrict the measurement exactly to the area covered in the irradiations with the $\alpha$ beam. While the statistical uncertainties were smaller than 0.3\%, the definition of the sample mass was affected by systematic uncertainties of 4.0\% and 4.6\% for the Mo and the Sn samples, respectively.

\subsection{Irradiations and $\gamma$ counting}

\begin{figure}[!htb]
\includegraphics{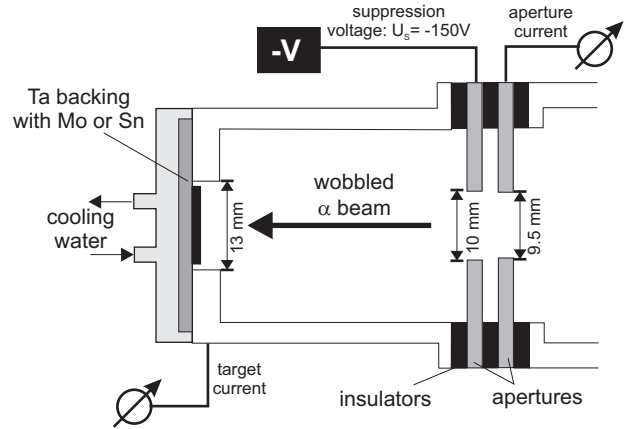}
\caption{\label{fig:PTB}Experimental setup at the PTB Braunschweig. }
\end{figure}

A sketch of the beamline setup at the PTB is shown in Fig.~\ref{fig:PTB}. The activation chamber is designed as a Faraday cup, and the charge deposited on the sample was recorded in small time steps by a current integrator for later correction of beam fluctuations. Secondary electrons were suppressed by a negatively charged (U$_S$=\mbox{-150~V}) diaphragm at the entrance of the activation chamber.  

For each energy step the $\alpha$-beam spot was adjusted by means of a quartz window so that the sample was illuminated as complete and homogeneously as possible by wobbling of the $\alpha$-beam. The energy of the $\alpha$ beam was defined within $\pm$25 keV uncertainty by means of the field calibration of two analyzing magnets as well as by a time-of-flight measurement of the particle velocity \cite{Boe02}. The effective $\alpha$-energy $E^{\rm{eff}}$ was calculated from the expression 4.99 in \cite{Ili07}

\begin{equation}
	E^{\rm{eff}}=E_{\rm{c.m.}}-\Delta+\Delta \cdot \left(-\frac{\sigma_2}{\sigma_1-\sigma_2}+ \left[\frac{\sigma_1^2 + \sigma_2^2}{2(\sigma_1-\sigma_2)^2}\right]^{1/2}  \right). \label{eq:eff}
\end{equation}

where $\Delta$ is the target thickness calculated with the Monte Carlo program SRIM 2003 \cite{SRIM03}, $E_{\rm{c.m.}}$ the respective center-of-mass energy, and $\sigma_1$, $\sigma_2$ the measured cross sections of two neighboring points. As can be seen from Eq.~\ref{eq:eff} the error bars of $E^{\rm{eff}}$ become asymmetric, with the smaller component in positive direction due to the correction factor in brackets (see Tables \ref{tab:92}, \ref{tab:94}, and \ref{tab:112}).

\begin{table*}[!htb]
\caption{\label{tab2} Decay properties of product nuclei and corrections for $\gamma$-ray counting}
\renewcommand{\arraystretch}{1.1} 
\begin{ruledtabular}
\begin{tabular}{ccccccc}
Product & $t_{1/2}$  	& $E_\gamma$ & $I_\gamma$ & Reference & \multicolumn{2}{c}{Summing corrections [\%]}  \\
nucleus &  						& [keV]    	&  per decay  & & $\gamma$ Cascades   &  X Rays       \\
\hline
$^{95}$Ru  & (1.643 $\pm$ 0.014) h & 336.4 & 0.702 $\pm$ 0.005   & \cite{nds95} & 7 $\pm$ 1 & 11.2 $\pm$ 0.4	\\
           &                     & 626.8 & 0.178 $\pm$ 0.005   & & 7 $\pm$ 1 &      ``        \\
\hline
$^{97}$Ru  & (2.9 $\pm$ 0.1) d     & 215.7 & 0.856 $\pm$ 0.013 \footnotemark & \cite{nds97} & $<$0.2 & 12.4 $\pm$ 1.5 \\
           &                     & 324.4 & 0.1079 $\pm$ 0.0017 & &    ``     &      ``        \\
\hline
$^{116}$Te & (2.49 $\pm$ 0.04) h   &  93.7 & 0.331 $\pm$ 0.021  & \cite{nds116} & $<$0.2 & 15.6 $\pm$ 1.7 \\
\end{tabular}
\end{ruledtabular}
\footnotetext[1]{No uncertainty given in literature. Assumed uncertainty 1.6\%, as for the 324.4~keV line.}
\end{table*}    
  
During the activations, the samples were irradiated with average $\alpha$-beam currents of 5 $\mu$A for periods between 20 minutes and 14 hours, depending on isotope and cross section. For the Sn samples the waiting time between the end of activation and the start of activity counting was extended to one hour to eliminate the $^{38}$K decay from the concomitant $^{35}$Cl($\alpha,n$)$^{38}$K reaction ($t_{1/2}$= 7.5~min). The induced $\gamma$ activities were measured with a calibrated HPGe detector. Natural backgrounds were suppressed by a shielding consisting of an inner layer of 5 mm copper surrounded by 10 cm of lead. The samples were analyzed as described 
in Ref. \cite{rapp02} using the $\gamma$-ray energies and intensities of the characteristic lines listed in Table \ref{tab2}. In cases, where two $\gamma$-ray lines could be used in data analysis, the results were averaged by weighting with the respective transition probabilities.

Summing corrections due to the coincident detection of $\gamma$-cascades were evaluated by means of simulations with the Monte Carlo code CASC \cite{Jaa93} using the decay schemes of the corresponding product nuclei $^{95}$Ru, $^{97}$Ru, and $^{116}$Te \cite{nds95,nds97,nds116}. The resulting corrections for $\gamma$ summing are given in Table \ref{tab2} together with the summing corrections due to the coincident detection of X rays. The analysis of the induced activity in the $^{112}$Sn samples via the 93.7 keV transition in the decay of $^{116}$Te was complicated by a constant background activity of (0.031$\pm$0.003) cts/s, corresponding presumably to the Po-K$_\beta$-line, as well as by the characteristic 93.3 keV $\gamma$ line from the decay of $^{67}$Ga. The latter background was produced by $^{63}$Cu($\alpha,\gamma$) reactions due to a copper contamination of the samples. The background component from the $^{67}$Ga decay could be corrected by means of the associated $^{67}$Ga line at 184.6 keV. 

\subsection{Uncertainties}
The systematic uncertainties (Table~\ref{tab3}) are governed by the target thickness and the detector efficiency. For $^{97}$Ru, the uncertainty in the half-life gives rise to an additional contribution of 3.4\%. The same holds for the $\gamma$ intensity of the 93.7~keV line from the decay of $^{116}$Te with an uncertainty of 6.3\% \cite{nds116}. The overall uncertainties are 5.3\% and 6.3\% for the $^{92}$Mo$(\alpha,n)$ and $^{94}$Mo$(\alpha,n)$, and 8.7\% for the $^{112}$Sn$(\alpha,\gamma)$$^{116}$Te cross sections.

\begin{table}[!htb]
\caption{Systematic uncertainties (in \%).}\label{tab3}
   \begin{ruledtabular}
\begin{tabular}{lccc}
Source of uncertainty & $^{92}$Mo($\alpha,n$) & $^{94}$Mo($\alpha,n$) & $^{112}$Sn($\alpha,\gamma$) \\
\hline 
Target thickness                   & 4.0  & 4.0 & 4.6  \\
Efficiency  											 & 3.0  & 3.0 & 3.0  \\ 
Beam current            					 & 1.0  & 1.0 & 1.0  \\ 
$\gamma$ intensity       					 & 0.7  & 0.5 & 6.3  \\ 
Decay constants                    & 0.9  & 3.4 & 1.6  \\
Summing corrections                & 1.1  & 1.5 & 1.7  \\
\hline
Overall uncertainty     					 & 5.3  & 6.3 & 8.7\footnotemark  \\
\end{tabular}
\end{ruledtabular}
\footnotetext[1]{Does not include effect of background line (see text).}
\end{table}

\section{Results and comparison with previous data}\label{results}
\begin{figure*}[!htb]
\includegraphics{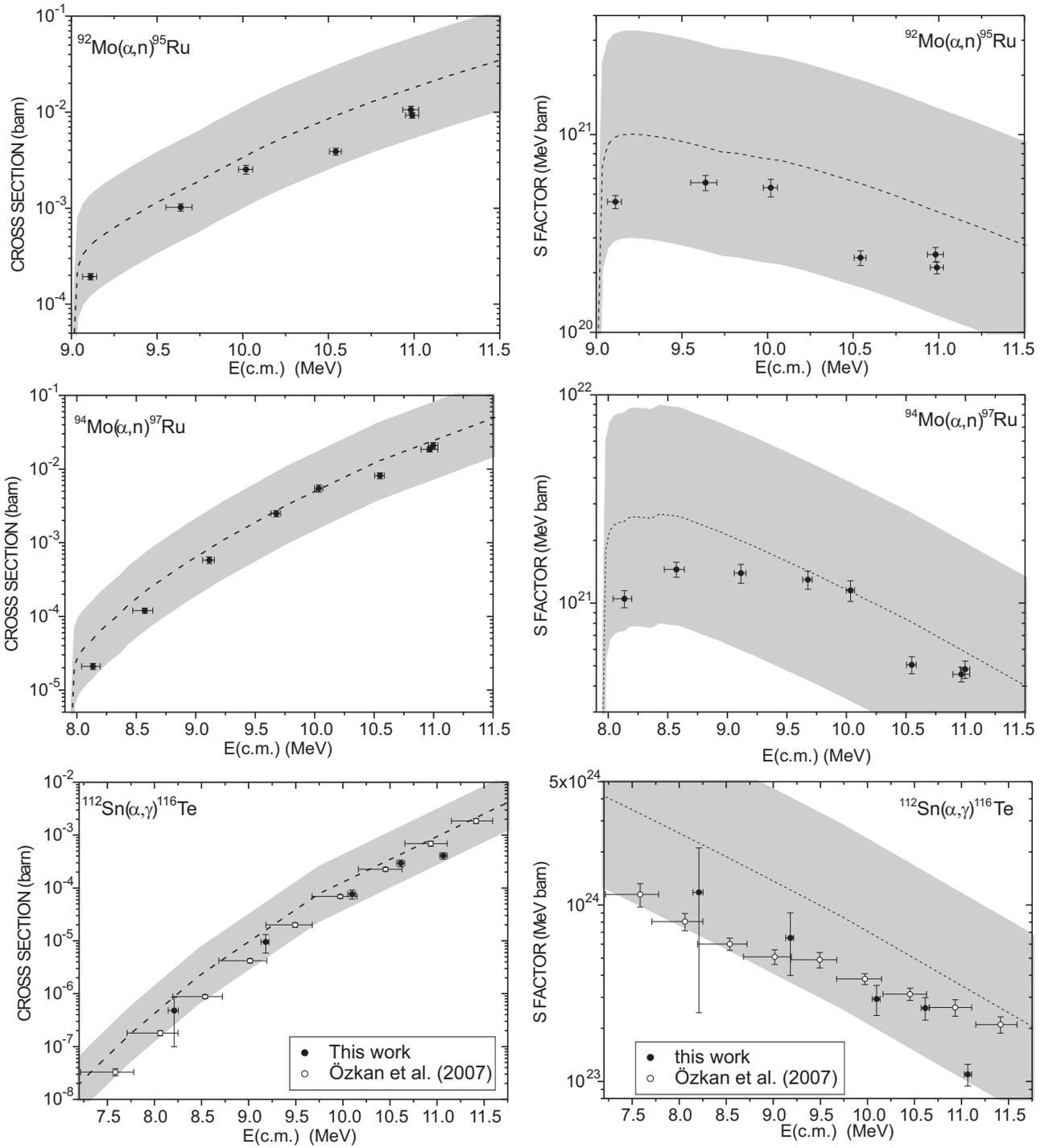}
\caption{\label{fig:plots}Cross section and $S$ factor for $^{92}$Mo$(\alpha,n)$$^{95}$Ru, $^{94}$Mo$(\alpha,n)$$^{97}$Ru, and $^{112}$Sn$(\alpha,\gamma)$$^{116}$Te in comparison with standard NON-SMOKER predictions (dashed line). The grey area shows the estimated uncertainty of the prediction (factor 10).}
\end{figure*}

The results are summarized in Tables \ref{tab:92}, \ref{tab:94}, and \ref{tab:112}, and in Fig.~\ref{fig:plots} in form of cross sections and astrophysical $S$ factors. The cross sections of all three reactions (Fig.~\ref{fig:plots}) are systematically lower by factors of two to three compared to the global Hauser-Feshbach predictions from NON-SMOKER (dashed line). 

For $^{112}$Sn($\alpha,\gamma$)$^{116}$Te our data points follow the Hauser-Feshbach dependence within the uncertainties, whereas the results from \"Ozkan et al. \cite{OEG07}, which were obtained with rather thick targets and represent averaged cross sections over energy bins of 440 to 565~keV, seem to exhibit a slightly different energy trend. Unfortunately, the present measurements at lower energies were severely hampered by an interfering background line from $^{63}$Cu$(\alpha,\gamma)$$^{67}$Ga reactions on the Cu impurity of the samples. This background could be only normalized by a much weaker $^{67}$Ga line at 184.6~keV, resulting in additional systematic and statistical uncertainties.

\begin{figure*}[!hp]
\includegraphics{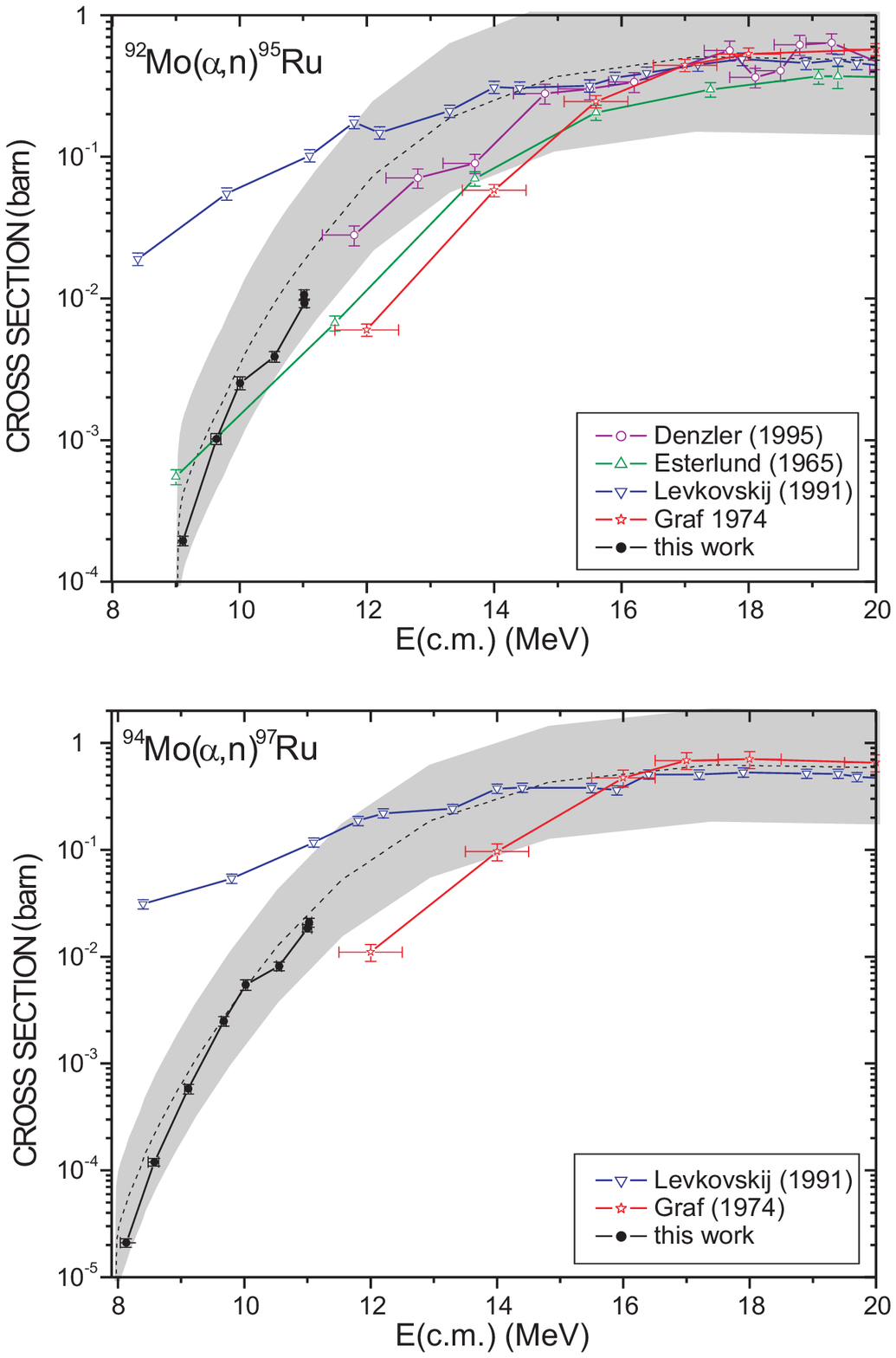}
\caption{\label{fig:comp-an}(Color online) Comparison of previous experiments at higher energies (up to 20 MeV) with our results and with the standard NON-SMOKER predictions (dashed line). The grey area shows the uncertainty of the prediction (factor 10).  }
\end{figure*}

For the $(\alpha,n)$ cross sections some older measurements were available for comparison (Fig.~\ref{fig:comp-an}). The data given by Levkovskij \cite{Lev91} is reproducing the NON-SMOKER cross sections in both cases perfectly for energies above 13~MeV, but there is a systematic deviation (compared to all other data) below this energy, presumably due to an undetected background component. This problem is particularly apparent at the lowest data point at 8.4~MeV, whereas the $^{92}$Mo$(\alpha,n)$$^{95}$Ru threshold is at 9.39~MeV. A similar problem exists for $^{94}$Mo$(\alpha,n)$.

The values from Esterlund et al. \cite{EP65} for $^{92}$Mo$(\alpha,n)$ and Graf et al. \cite{GM74} for $^{92,94}$Mo$(\alpha,n)$ seem to be low compared to the energy trend below 15~MeV, but agree like the Levkovskij data rather well with the predictions at higher energies. The only data which follow the Hauser-Feshbach trend also at lower energies down to 11.8~MeV and at the same time fits well with our highest data point are the $^{92}$Mo$(\alpha,n)$ cross sections of Denzler et al. \cite{DRQ95}. This comparison shows the necessity to measure $(\alpha,n)$ cross sections -- especially for $p$-process studies -- down to the threshold.

\begin{table}[!htb]
\caption{\label{tab:92} Cross sections and $S$ factors for $^{92}$Mo$(\alpha,n)$$^{95}$Ru.}
\renewcommand{\arraystretch}{1.2} 
\begin{ruledtabular}
\begin{tabular}{ccc}
$E{\rm ^{eff}_{c.m.}}$ & Cross section  & $S$ factor \\
$$ [MeV]  & [mbarn] & $\times$10$^{20}$[MeV barn] \\
\hline
10.989 $^{+0.038}_{-0.039}$ & 9.26 $\pm$ 0.65			& 2.13 $\pm$ 0.15 \\
10.981 $^{+0.046}_{-0.047}$ & 10.6 $\pm$ 0.9 			& 2.47 $\pm$ 0.21 \\
10.543 $^{+0.033}_{-0.037}$ & 3.88 $\pm$ 0.33 	  & 2.38 $\pm$ 0.20 \\
10.019 $^{+0.039}_{-0.043}$ & 2.53 $\pm$ 0.26 	  & 5.39 $\pm$ 0.55 \\
 9.638 $^{+0.066}_{-0.087}$ & 1.02 $\pm$ 0.09 	  & 5.71 $\pm$ 0.50 \\
 9.112 $^{+0.036}_{-0.047}$ & 0.194 $\pm$ 0.015  	& 4.57 $\pm$ 0.35 \\
\end{tabular}
\end{ruledtabular}
\end{table}

\begin{table}[!htb]
\caption{\label{tab:94} Cross sections and $S$ factors for $^{94}$Mo$(\alpha,n)$$^{97}$Ru.}
\renewcommand{\arraystretch}{1.2} 
\begin{ruledtabular}
\begin{tabular}{ccc}
$E{\rm ^{eff}_{c.m.}}$ & Cross section  & $S$ factor \\
$$ [MeV]  & [mbarn] & $\times$10$^{20}$[MeV barn] \\
\hline
10.997 $^{+0.039}_{-0.040}$  & 20.90 $\pm$ 2.00 	  & 4.81 $\pm$ 0.46 \\
10.966 $^{+0.070}_{-0.073}$  & 18.50 $\pm$ 1.50 	  & 4.56 $\pm$ 0.37 \\
10.548 $^{+0.038}_{-0.044}$  & 8.14 $\pm$ 0.77 	  	& 5.06 $\pm$ 0.48 \\
10.034 $^{+0.034}_{-0.036}$  & 5.47 $\pm$ 0.62 	  	& 11.50 $\pm$ 1.30 \\
9.676 $^{+0.037}_{-0.043}$   & 2.49 $\pm$ 0.25 	  	& 12.96 $\pm$ 1.30 \\
9.114 $^{+0.042}_{-0.056}$   & 0.58 $\pm$ 0.06 	  	& 13.92 $\pm$ 1.44 \\
8.572 $^{+0.067}_{-0.102}$   & 0.12 $\pm$ 0.01 	  	& 14.52 $\pm$ 1.21 \\
8.136 $^{+0.061}_{-0.095}$   & 0.021 $\pm$ 0.002   	& 10.49 $\pm$ 1.00 \\
\end{tabular}
\end{ruledtabular}
\end{table}

\begin{table}[!htb]
\caption{\label{tab:112} Cross sections and $S$ factors for $^{112}$Sn$(\alpha,\gamma)$$^{116}$Te.}
\renewcommand{\arraystretch}{1.2} 
\begin{ruledtabular}
\begin{tabular}{ccc}
$E{\rm ^{eff}_{c.m.}}$ & Cross section  & $S$ factor \\
$$ [MeV]  & [$\mu$barn] & $\times$10$^{23}$[MeV barn] \\
\hline
11.067 $^{+0.042}_{-0.038}$ & 408 $\pm$ 57 			& 1.10 $\pm$ 0.15 \\
10.615 $^{+0.040}_{-0.043}$ & 295 $\pm$ 43 			& 2.61 $\pm$ 0.38 \\
10.097 $^{+0.037}_{-0.045}$ & 77 $\pm$ 15 		  & 2.93 $\pm$ 0.57 \\
9.180 $^{+0.036}_{-0.048}$  & 9.52 $\pm$ 3.68 	& 6.51 $\pm$ 2.52 \\
8.209 $^{+0.043}_{-0.064}$  & 0.48 $\pm$ 0.38 	& 11.78 $\pm$ 9.33 \\
\end{tabular}
\end{ruledtabular}
\end{table}

\section{Test of the optical $\alpha$+nucleus potential}\label{sec:alphapot}
The present experimental $S$ factors were compared to NON-SMOKER Hauser-Feshbach calculations employing different optical $\alpha$+nucleus potentials while keeping the standard inputs for the other parameters. The standard optical potential used in NON-SMOKER \cite{rath00} is the one by \mbox{McFadden} and Satchler \cite{McS66}, derived from scattering data above 60 MeV. This is also the potential used in the standard NON-SMOKER predictions shown in Figs.\ \ref{fig:plots} and \ref{fig:comp-an}. Another global potential is the one by England et al. \cite{EBN82}, also fitted to data at higher energies. A more recent global potential, which also included reaction data, was given by Avrigeanu et al. \cite{AOP03}. Finally, Fr\"ohlich et al. (see \cite{RF03}, erratum in \cite{RFCor03}) used reaction data at $A\approx 150$ to derive a low-energy potential which also reproduces data at lower masses (e.g., \cite{RF03,GKE06}).

\begin{figure*}[!hp]
\includegraphics{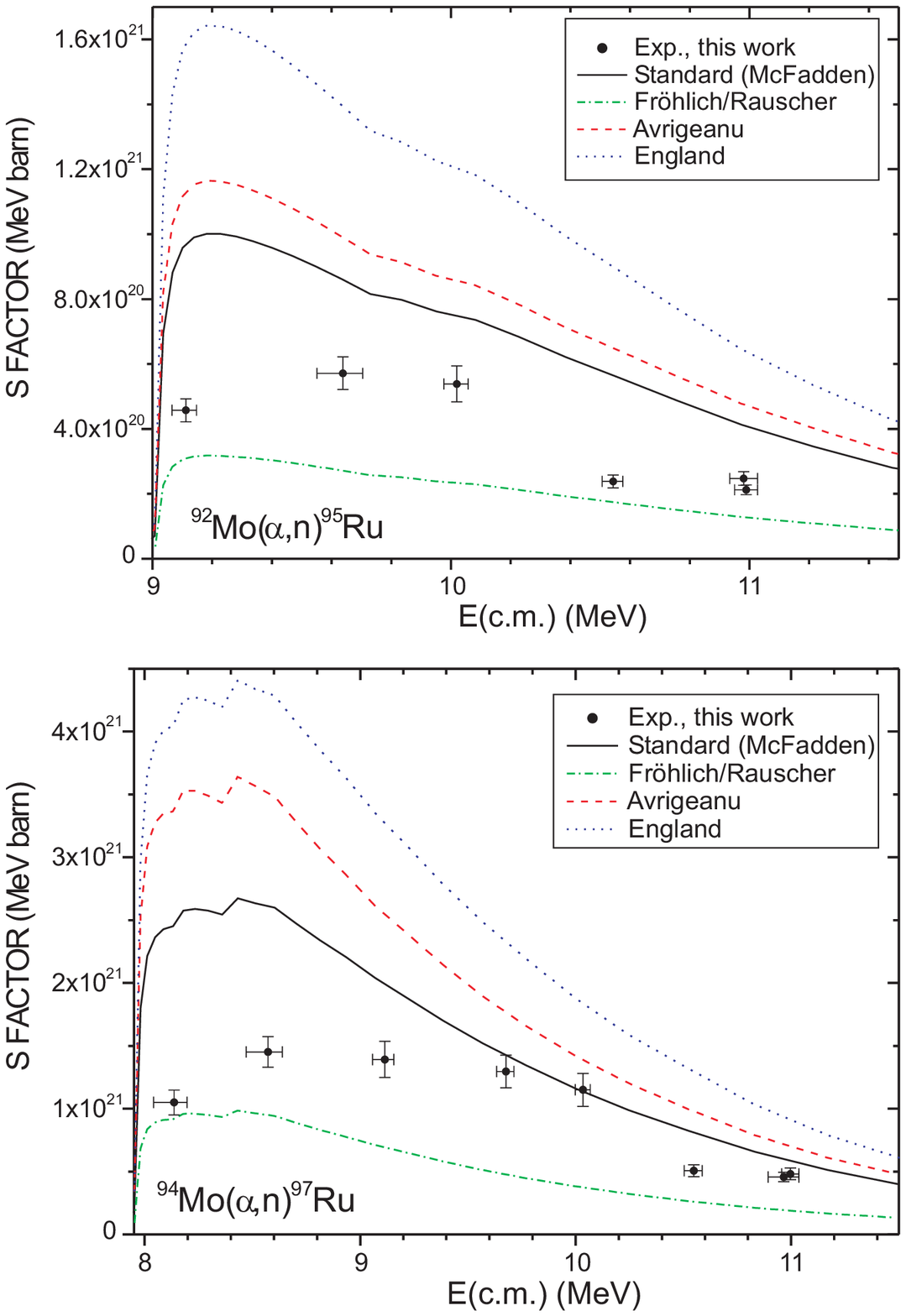}
\caption{\label{fig:potcomp-9294}(Color online) Comparison of $S$ factors obtained with different optical potentials in the NON-SMOKER code for the reactions $^{92}$Mo$(\alpha,n)$$^{95}$Ru and  $^{94}$Mo$(\alpha,n)$$^{97}$Ru (see text for details).}
\end{figure*}

The reaction $^{92}$Mo$(\alpha$,n)$^{95}$Ru is mainly sensitive to the optical $\alpha$ potential in the measured energy range. Only up to a few keV above the threshold, the averaged neutron widths are smaller than the $\alpha$ widths and therefore the sensitivity is higher to the neutron widths close to the threshold.  As can be seen in Fig.~\ref{fig:potcomp-9294}, the theoretical results obtained with the different optical potentials display the same energy behavior but with different absolute values. This energy dependence describes that of the data reasonably well. The $S$ factors obtained with the McFadden and Satchler potential have to be divided by 1.8, whereas the ones obtained with the Fr\"ohlich potential are a factor of 1.8 too low. All other potentials used lead to larger deviations. An optical potential for $^{92}$Mo+$\alpha$ was derived from scattering data and carefully extrapolated to low energy in Ref. \cite{FGM03}. We do not show the results with this potential because they coincide with the Avrigeanu et al. results at the low energy range shown here, whereas they coincide with the McFadden and Satchler $S$ factors at the high end of the energy range. Thus, its energy dependence is slightly different from that of the other potentials and the data.

Also the reaction $^{94}$Mo$(\alpha,n)$$^{97}$Ru is almost exclusively sensitive to the $\alpha$ widths and shows only a small dependence on the neutron widths, except close to the threshold. Inspection of Fig.~\ref{fig:potcomp-9294} evidently shows that the energy dependence of the data cannot be reproduced by any of the potentials tried here. The peak $S$ factor is shifted to higher energy in the data as compared to the one found in the calculated $S$ factor. The cause of this remains unclear. The importance of the neutron widths does not extend beyond a few tens of keV above the ($\alpha$,n) threshold. The impact of width fluctuation corrections is also confined to a small window above the reaction threshold. Similar to the case above, multiplication of the results obtained with the Fr\"ohlich potential with a factor 1.8 and division of the ones obtained with the McFadden and Satchler potential by 1.7 lead to an improved reproduction of the data. However, because of the distinctly different energy dependence none of these solutions are satisfactory.

\begin{figure*}[!htb]
\includegraphics{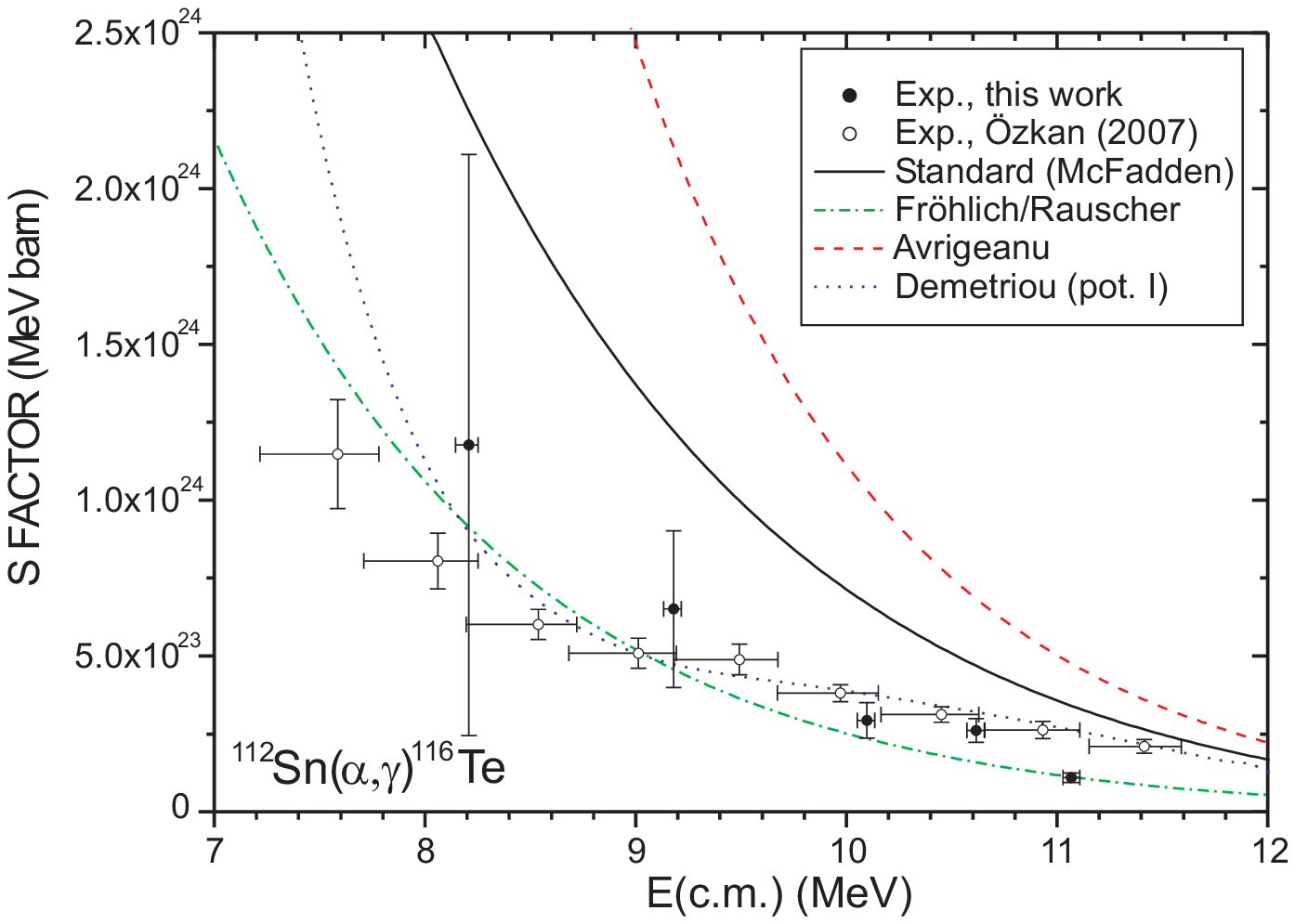}
\caption{\label{fig:potcomp-112}(Color online) Comparison of $S$ factors obtained with different optical potentials in the NON-SMOKER code for the reaction $^{112}$Sn$(\alpha,\gamma)$$^{116}$Te (see text for details).}
\end{figure*}

It may be surprising at first glance that the $S$ factors of the reaction $^{112}$Sn$(\alpha,\gamma)$$^{116}$Te are much more sensitive to the $\alpha$ widths than to the $\gamma$ widths. However, due to the Coulomb barrier, the $\alpha$ widths are smaller by one to several orders of magnitude than the $\gamma$ widths in the energy range covered in the measurement. There are two data sets available, the one obtained in this work and the one by \"Ozkan et al. \cite{OEG07}. Both of them are compared to theoretical results in Fig.~\ref{fig:potcomp-112}. Concerning the theoretical predictions using different optical $\alpha$ potentials it is not surprising that the \"Ozkan et al. data at the high energy range are best described employing the potential I of Demetriou et al. \cite{DGG02}. A preliminary version of these data was part of the procedure to derive the Demetriou et al.\ potential, a global potential for heavy targets. Similar agreement was already found in a study of $\alpha$ scattering on $^{112}$Sn \cite{GFG05}. In that study, an optical potential was derived from scattering data and then used to compute the $^{112}$Sn$(\alpha,\gamma)$$^{116}$Te $S$ factors. Peculiarly, however, it was found that the derived potential did not reproduce the \"Ozkan et al. reaction data well. Much better agreement was achieved with the Demetriou et al.\ potential and best agreement was obtained when using the Fr\"ohlich potential, although the latter did not describe the scattering data. Here, we confirm these findings \cite{GFG05}. We do not show the result with the optical potential derived in \cite{GFG05} separately because it basically coincides with the one obtained with the McFadden and Satchler potential. 

Interestingly, our new experimental data seem to indicate a slightly steeper energy dependence than the \"Ozkan data. This yields good agreement (within the error bars) with the prediction making use of the Fr\"ohlich-Rauscher potential (with exception of the data point at 10.615~MeV). The calculation using the Demetriou potential \cite{DGG02} exceeds our data (including error bar) at the three highest energies and is compatible to our data only at the two lowest energies. Comparing to the \"Ozkan et al. data, the Fr\"ohlich-Rauscher potential gives acceptable agreement only below 9.5 MeV and underestimates the $S$ factors above that energy. On the other hand, the Demetriou et al. potential excellently reproduces the \"Ozkan et al. data above 8.5 MeV but strongly overestimates them below that energy. 

\section{Summary}
We have measured the $(\alpha,n)$ cross sections of the $p$ nuclei $^{92,94}$Mo and the $(\alpha,\gamma)$ cross section of the lightest tin isotope $^{112}$Sn. For $^{94}$Mo$(\alpha,n)$ no data for comparison at lower energies was available, but for $^{92}$Mo our energy trend seems to fit well with the results of Denzler et al. \cite{DRQ95} starting at 12 MeV. For $^{112}$Sn$(\alpha,\gamma)$$^{116}$Te we can essentially reproduce the thick-target cross sections from \"Ozkan et al. \cite{OEG07} although we find slightly lower values above 10 MeV. Additionally, we studied the sensitivity of Hauser-Feshbach predictions performed with the code NON-SMOKER for these reactions. We find that the cross sections and $S$ factors are mostly sensitive to the $\alpha$ widths and thus to the optical $\alpha$+nucleus potential employed in the calculations. The standard, global predictions -- utilizing the global potential by McFadden-Satchler \cite{McS66} -- systematically overestimate the data by a factor of two. The potential of Fr\"ohlich and Rauscher \cite{RF03,RFCor03} describes well the reaction data of $^{112}$Sn$(\alpha,\gamma)$$^{116}$Te but underestimates the $^{92,94}$Mo data by a factor of about two. The energy dependence of $^{94}$Mo($\alpha$,n)$^{97}$Ru cannot be reproduced by the statistical model calculations.

\begin{acknowledgments}
We thank the operating team O. D\"ohr, H. Eggestein, T. Heldt, and M. Hoffmann for providing the $\alpha$-beam at the PTB cyclotron.
This work was partially supported by the Swiss NSF (grant 2000-113984/1).
\end{acknowledgments}

\newpage 

\end{document}